\documentstyle[sprocl,epsfig]{article}

\def\be{\begin{equation}}
\def\ee{\end{equation}}
\def\bea{\begin{eqnarray}}
\def\eea{\end{eqnarray}}

\newcommand{\equ}[1]{Eq.\,(\ref{#1})}
\newcommand{\eqs}[1]{Eqs.\,(\ref{#1})}

\newcommand{\ew}{electroweak~}

\newcommand{\gsim}{\;\rlap{\lower 3.5 pt \hbox{$\mathchar \sim$}} \raise 1pt
 \hbox {$>$}\;}
\newcommand{\lsim}{\;\rlap{\lower 3.5 pt \hbox{$\mathchar \sim$}} \raise 1pt
 \hbox {$<$}\;}
\newcommand{\smallz}{{\scriptscriptstyle Z}} 
\newcommand{\smallw}{{\scriptscriptstyle W}} %
\newcommand{\smallh}{{\scriptscriptstyle H}} %

\newcommand{\mz}{M_\smallz}
\newcommand{\mw}{M_\smallw}
\newcommand{\mh}{M_\smallh}
\newcommand{\mt}{M_t}
\def\pl#1#2#3{{\it Phys. Lett. }{\bf B#1~}(19#2)~#3}
\def\zp#1#2#3{{\it Z. Phys. }{\bf C#1~}(19#2)~#3}
\def\prl#1#2#3{{\it Phys. Rev. Lett. }{\bf #1~}(19#2)~#3}

\def\pr#1#2#3{{\em Phys. Rev. }{\bf D#1~}(19#2)~#3}
\def\np#1#2#3{{\em Nucl. Phys. }{\bf B#1~}(19#2)~#3}

\newcommand{\msbar}{\overline{\rm MS}}
\newcommand{\seff}{\sin^2\theta_{eff}^{lept}}
\newcommand{\as}{\alpha_s}

\begin{document}

\title{PRECISION TESTS OF THE STANDARD MODEL: HIGHER
ORDER CORRECTIONS AND THE HIGGS MASS\footnote{Talk presented at RADCOR98,
{\it International Symposium on Radiative Corrections},
  Barcelona, September 1998.}}

\author{P. GAMBINO}

\address{Technische Universit\"at M\"unchen,
James Franck Str.,\\
Garching, D-85748 Germany\\
E-mail: gambino@physik.tu-muenchen.de}


\maketitle\abstracts{ Recent 
calculations of the two-loop electroweak effects enhanced by powers
of the top mass have been implemented in the main electroweak libraries
and have an important effect on the indirect determination of the
Higgs mass $\mh$. I  briefly review the main results of these calculations
and discuss in detail  the residual uncertainties and their impact on 
the global fit to $\mh$. The perspectives for the near future
 are also considered. }


The overall agreement of the present precision data with the Standard Model
(SM) is quite good \cite{teubert}. The value of $\mt$ estimated by
a global fit, for example, is $\mt=161.1^{+8.2}_{-7.1}$ GeV, which compares 
well to the direct determination of $\mt$ at  the Tevatron.
One can  try to obtain similar indirect constraints on the mass of the
Higgs boson (see Fig.~1). 
However, the sensitivity of the various precision observables to 
$\mh$ is much milder than the one to the top mass. The extraction of the
relevant information is therefore more difficult and 
delicate in this case, and requires a
careful  consideration of some two and three-loop effects and of the residual
errors of the theoretical predictions.

Before going into details,
it may be interesting to get a quick idea of the size and the importance of 
genuine two and three-loop corrections when we calculate the main precision
observables.  Table 1 shows the shifts induced by the known QCD and \ew
higher order effects on the one-loop prediction of $\mw$ and $\sin^2
\theta_{eff}^{lept}$ for a few values of $\mh$.  
The two-loop $O(\alpha_s)$ corrections have been
calculated in \cite{QCD2}, the three-loop  $O(\alpha_s^2)$ related to the
leading top contribution in \cite{QCD3a}, and 
the heavy top expansion of the complete $O(\alpha_s^2)$ effect 
in \cite{QCD3b}. For what concerns the purely 
\ew irreducible two-loop effects, only the first term 
\cite{barb}, $O(g^4 \mt^4/\mw^4)$, and second  
term \cite{dgv,dgs,dg}, $O(g^4 \mt^2/\mw^2)$, 
of an expansion in  powers of the top mass are known. 
In particular, the second term 
is scheme dependent and the results in Table 1 refer
to  the  $\msbar$ scheme of Ref. \cite{msbar}.

It is remarkable that all the effects listed in Table 1 have the same 
sign, corresponding  to a {\it screening} of the leading one-loop
contribution, $O(G_\mu \mt^2)$, 
which is due to the non-decoupling top quark effect in  $\Delta\rho$. 
Without these higher order corrections 
the precision tests of the SM would point
towards a much  lighter top quark, about 20 GeV less, and 
would be in conflict with the Tevatron measurement of $\mt$. 
In fact,  the bulk of these effects is directly  connected to the top quark.
Conversely, one may also note 
that  their total  is about 100 MeV and 5.5 $10^{-4}$
for $\mw$ and $\sin^2\theta_{eff}^{lept}$, respectively.  This  has to be
compared with the present experimental accuracy~\cite{teubert}, 
of about 60 MeV and 2 $10^{-4}$. Notice that the enhancement of the 
{\it screening} due to the $O(g^4 \mt^2/\mw^2)$ 
correction is not only a feature of the $\msbar$ scheme, since the trend is 
more or less common to all the popular schemes. 
This will be more clear in the following.

\renewcommand{\arraystretch}{1.2}
\begin{table}[t] 
\begin{tabular}{|c||c||c|c|c|c|c|}\hline
$\mh$  & $\mw$& $O(\alpha\alpha_s)$  &$O(\alpha\alpha_s^2 )$   & 
$O(g^4 m_t^4)$& $O(g^4 m_t^2)$
&   total\\  \hline\hline
100 &80.480 & -63 & -12 & -10 & -11 & -96 \\ \hline
300 & 80.409 & -63 & -12 & -16 &  -9 & -100   \\ \hline
800 & 80.332 &  -63 & -12 &  -20& -7 & -102\\ \hline
\end{tabular}            
\end{table}
\begin{table}[t] 
\begin{tabular}{|c||c||c|c|c|c|c|}\hline
$\mh$ &$\sin^2 \theta_{eff}^{lept}$ 
& $O(\alpha\alpha_s)$  &$O(\alpha\alpha_s^2)$   & 
$ O(g^4 m_t^4)$& $O(g^4  m_t^2)$&  total\\  
\hline\hline
 100& 0.23108 & 3.5& 0.6 &0.5  & 0.6& 5.2 \\ \hline 
 300& 0.23154 & 3.5 &0.6  &0.9 & 0.6  & 5.6    \\ \hline
 800& 0.23207 & 3.5 & 0.6 & 1.2 & 0.4 & 5.7\\ \hline
\end{tabular}            
\caption{One-loop  predictions of $\mw$ and of the effective sine with
corresponding  shifts (in MeV and $10^{-4}$)
induced by known two and three-loop  effects. The numbers refer to the case 
of the $\msbar$ scheme  for
$\mt=175$\,GeV and  $\alpha_s(\mz)=0.119$. $\mw$ and $\mh$ are 
expressed in GeV.}
\end{table}
What is the effect of all this on the indirect determination of $\mh$? 
From Table 1 we  see that among these higher order effects
 only the purely \ew  corrections slightly
modify the one-loop $\mh$ slope of the predicted $\mw$ and effective sine.
 In other schemes their impact, because of  a rearrangement of the
reducible contributions, may be larger.
Most of the  influence of higher order
effects on the Higgs mass  determination  is however indirect, through i) the 
{\it screening} of the Veltman correction which gives the bulk of the \ew
one-loop radiative corrections ($\mt$ and $\mh$ are indeed strongly correlated 
in the global fit, as we will see); ii) the reduction of the theoretical error
that we expect when we include new higher order corrections.
This last point can be visualized by the reduction in the size of the 
blue band in the $\chi^2$ vs $\mh$ plot of the global fit prepared by 
the LEP-SLD Electroweak Working Group (EWWG)~\cite{teubert} (see Fig.~1), with
respect to the same plot before the implementation of the latest \ew 
higher order corrections.

Another topic I would like to briefly touch before expanding on the main
subject of this talk is related to something mentioned in the two nice review
talks given at this conference by F. Teubert~\cite{teubert} and
W. Hollik~\cite{hollik}. Unlike  a few years
ago, we have now a very strong experimental evidence for \ew radiative
corrections beyond the (trivial) running of the electromagnetic coupling
contained in $\Delta \alpha$.\cite{teubert} 
If we argue that, after the discovery of the top
quark, the fermionic sector of the SM is well-established and that its
couplings with the vector bosons have been tested in many experiments, we can
wonder what is the evidence for the purely bosonic contributions
that contain virtual bosons and
represent the core of the gauge structure and of the Higgs mechanism of the SM.
They involve all the tri-linear couplings as well as the Higgs boson and
represent a gauge-invariant subset of contributions which is unambiguous but
numerically subleading. 
Already a few years ago\cite{evid}, it was possible to show that the single
measurement of the effective sine combined with the lower bound on the top mass
$\mt>131$ GeV provided evidence for the bosonic corrections of the theory at 
the level of $4\sigma$ (see also \cite{bosonic}). 
Today the same analysis, based on a conservative 
limit $\mt>164$ GeV and on the present measurement of $\seff$ which is 
twice as accurate as it was at the time of ~\cite{evid}, 
establishes the necessity of purely bosonic contributions at the level of 
8.9$\sigma$.

\begin{figure}[t]
\centering
\mbox{
\epsfig{file=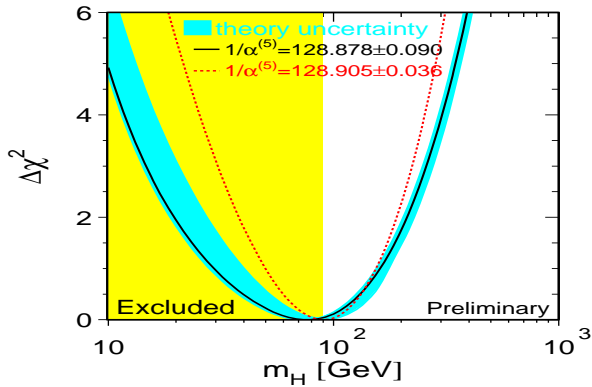 
        ,height=6cm  
        ,width=8cm   
}}
\caption{The latest global $\mh$ fit  prepared 
by the LEP-SLD Electroweak Working Group.}
\end{figure}

As my  aim is to investigate the indirect Higgs boson mass determination, 
it will be sufficient to consider the three most precisely measured
quantities:
$\seff$, $\mw$, and $\Gamma_{l}$, the leptonic partial width of the $Z^0$
boson. All one-loop effects as well as some higher order effects have
extensively been studied in \cite{YB} (see \ \cite{hollik} for a comprehensive
 list of references). To go beyond that, the strategy is obviously to look for 
possible large higher order effects, and corrections enhanced by powers 
of heavy masses are prime candidates. In particular, it is possible to organize
the two-loop \ew corrections to the various observables in asymptotic series 
of the  heavy top mass 
and retain only the first two terms, $O(g^4 \mt^4/\mw^4)$ and
$O(g^4 \mt^2/\mw^2)$. This is suggested by the dominance of the top quadratic
contribution among \ew effects at the one-loop level.
In practice, one needs  to calculate at this order the  Thomson scattering,
the $W$ and $Z$ propagators, the muon decay, and the leptonic $Z^0$ decay.
Some details of these calculations are given in \cite{dgv,dgs,zako} and 
 more will appear in \cite{dg}.
The calculation has been performed in three different \ew schemes:
the $\msbar$ scheme of ~\cite{msbar}, and two very different implementation of
the {\it on-shell} scheme of \cite{si80}, which are defined in \cite{dgs}.
The results of the precise determination of $\seff$ and $\mw$ 
obtained by incorporating these and all other known two and three-loop 
radiative corrections 
are shown  in Tables 2 and 3. 
\begin{table}[t]
\[
\begin{array}{|c|  c c c c| c c c c|}\hline
M_H  & 
 {\rm OSI} & {\rm OSII} 
& \rm{\overline{MS}} & ZFitter & 
 {\rm OSI} & {\rm OSII} 
& \rm{\overline{MS}} & ZFitter
  \\  \hline\hline
65  &  .23131 & .23111 & .23122 &.23116 &
 80.411 & 80.422 & 80.420 & 80.420   \\ 
300  & .23212 & .23203 & .23203 & .23197 & 80.312 & 80.316 
& 80.319 & 80.320\\
1000 & .23280 & .23282 & .23272 & .23264  & 80.215 & 80.213 & 80.221 & 80.224
\\ \hline 
\end{array}            
\]\caption{Comparison of the predictions for $\seff$ and for $\mw$ (in GeV) in 
the three different schemes of \protect\cite{dgs} and by 
ZFitter\protect\cite{zfitter} before the inclusion of
the $O(g^4 \mt^2/\mw^2)$ corrections. $\mt=175$ GeV. }
\label{tab2}
\end{table}
\begin{table}[t]
\[
\begin{array}{|c|  c c c|c c c|}\hline
M_H  &  {\rm OSI} & {\rm OSII} 
& \rm{\overline{MS}} 
 &  {\rm OSI} & {\rm OSII} 
& \rm{\overline{MS}} \\  \hline\hline
65  &  .23132 & .23132 & .23130 &  80.404 & 80.404 & 80.406    \\ 
300  & .23209 & .23212 & .23209  & 80.308 & 80.307 & 80.309\\ 
1000 & .23275 & .23277 & .23275 & 80.216 & 80.215 & 80.216 \\ \hline 
\end{array}            
\]\caption{Same as in Table \ref{tab2} but  after the inclusion of
the $O(g^4 \mt^2/\mw^2)$ corrections. }
\end{table}
Table 2 reports the results before the implementation of the $O(g^4 \mt^2/\mw^2)$
effects, while  Table 3 shows the results including them.
We observe that i) the predictions of $\seff$ ($\mw$) in Table 3 are 
generally higher (lower)
than in Table 2, which corresponds to an enhancement of the screening of the
leading top contribution;
ii) the scheme ambiguity of the predictions is much smaller in Table
3 than in Table 2. This is consistent with the notion that the 
different schemes are equivalent up
to $O(g^4)$ corrections not enhanced by powers of heavy masses or by large 
logarithms, and that the residual corrections should be correspondingly
suppressed. 

The residual uncertainty due to uncalculated higher order effects 
of \ew origin
can be estimated by the scheme dependence and by the scale dependence in 
the $\msbar$ scheme.
In \cite{YB} the comparison of five different implementations of one-loop and
leading higher order corrections led to predictions of $\seff$ and $\mw$ which
differed 
by at most 2.8 $10^{-4}$ and 32 MeV, respectively. This is reflected
 in Table 2, where the maximum spreads are 2.0
$10^{-4}$ and 11 MeV. Judging by the results in Table 3, these scheme 
ambiguities have provided a realistic estimate of 
the $O(g^4 \mt^2/\mw^2)$ effects. Indeed, if we assign 
to each value in Table 2 an error equal to the scheme ambiguity 
for that observable,
all  the results in Table 3 are  within the range defined by 
this error, or close to it.   This method can
obviously  provide only  an order of magnitude estimate of the uncalculated
effects, but it gives a clear
 indication that the residual $O(g^4) $ corrections
should be quite smaller than the previous term of the heavy top expansion.
Using this criterion in Table 3 (and in 
the more complete tables of \cite{dgs}),
 we obtain $\delta \seff=\pm 4 \ 10^{-5}$ and $\delta\mw=\pm 2$ MeV, which can
 be considered as estimates of the uncalculated higher order \ew effects.
The scale dependence of the $\msbar$ results leads to  analogous 
estimates.\cite{dgps}

It remains to estimate the uncertainty due to higher order QCD corrections.
For the three observables we are considering, only gluonic
corrections to quark loops are involved.
Fortunately, a lot of work has been done a few years ago, and  analyses
based on very different
methods have led to consistent  results.\cite{qcderror} 
The largest part  of the QCD corrections concerns the 
leading $O(G_\mu \mt^2)$ one-loop contribution to $\Delta\rho$.
These corrections are quite large when $\Delta\rho$ is expressed in terms
of the pole mass of the top. Because of its long-distance sensitivity,
the latter is not a good expansion parameter and induces a QCD expansion
plagued by large and rapidly growing coefficients. 
A high-scale mass definition,
like the $\msbar$ one, implies a QCD series with much smaller coefficients
and presumably smaller higher corrections.
If  we adopt the estimate of the first reference of ~\cite{qcderror} for the
error on $\Delta\rho$, we 
obtain $\delta\mw\approx \pm3$ MeV and $\delta \seff\approx\pm$ 2 10$^{-5}$.
As there are additional QCD contributions, we may enlarge the theoretical error
to $\delta\mw=\pm5$ MeV and $\delta \seff=\pm$ 3 10$^{-5}$.

\renewcommand{\arraystretch}{1.3}
\begin{table}[t]
\[
\begin{array}{|c|  c| c | c |}\hline
{\rm source \ of \ uncertainty}
  & \delta\mw &\delta s^2_{eff} & \delta\Gamma_{lept}\\
 & \,({\rm MeV}) \,& \,(10^{-5})\, & \,({\rm KeV})\, \\\hline \hline
\delta\mt=5\ {\rm GeV} & 30  & 15  & 45 \\ 
 \delta\,\as(\mz)=0.003 & 2 & 1 & 3   \\ 
 \delta\mz=2 \ {\rm MeV} & 2 & 2 & 6  \\ 
\delta\alpha(\mz)/\alpha(\mz)=7 \ 10^{-4} & 13 & 23 & 12  \\ 
{\mh=65\,-\, 1000 {\rm GeV}} & 190 & 
140  & 240 \\
{\rm higher\ order\ EW} & 2 \,(11)^* & 4 \,(21)^* & 5 \,(18)^*  \\ 
{\rm higher\ order\ QCD} & 5  & 3 & 7  \\ \hline
\end{array}            
\]
\caption{Present parametric and intrinsic uncertainties in the calculation of
$\mw$, $\seff$, and $\Gamma_{lept}$. The numbers in the sixth row 
marked (unmarked) by $^*$ refer to the  
scheme/scale ambiguity of the calculation before (after)
 the implementation of the 
$O(g^4 \mt^2/\mw^2)$ corrections.}
\end{table}
Table 4 summarizes the various parametric and intrinsic errors 
in the calculation of the three observables. It should be noted that 
in all three cases the uncertainty induced by the unknown Higgs mass
is much larger than the experimental error, despite the fact that their 
$\mh$ dependence is in all cases only logarithmic.
Higher order \ew uncertainties are at the level of the error induced by 
the experimental measurement  of $\mz$.

The results of the calculation of the $O(g^4 \mt^2/\mw^2)$ effects on
$\mw$,\cite{dgv} $\seff$,\cite{dgs}  and $\Gamma_f$, 
for $f\neq b$,~\cite{dg} have been now
implemented, together with other new results less important for the $\mh$
determination,  in the latest versions of {\it TOPAZ0}~\cite{topaz0}  and 
{\it ZFitter}~\cite{zfitter}, which are routinely used for the global fits
by the EWWG. The numerical results of the two codes 
are in very good agreement with 
the ones of \ \cite{dgs} and among themselves.\cite{bardin}
The $\msbar$ scheme results have also been implemented in \cite{erler}, and
again there is good  agreement.

\begin{figure}[t]
\centerline{
\mbox{%
\epsfig{file=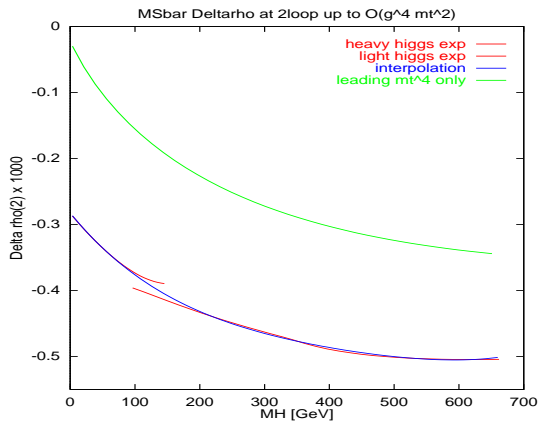 
        ,height=5.8cm  
        ,width=8cm   
}}}
\caption{$\mh$ dependence of the two-loop \ew corrections to 
$\Delta\hat{\rho}$ in the $\msbar$ scheme. The
result of the heavy top expansion up to its 
first (second) term is shown in the upper (lower) curve.
} \label{drho}
\end{figure}

A crucial question is now 
whether the approximation  based on the Heavy Top
Expansion (HTE) described above and now used for the fits is reliable.
The second term $O(g^4 \mt^2/\mw^2)$ of the two-loop HTE seems to be
quite important wrt the first (see Fig.\ref{drho} and Table 1), 
so the convergence of the HTE may be legitimately questioned. 
An important point to take into account in this
respect is that this is true mainly for a light Higgs,
 where the approximation
of keeping $\mw=0$ and $\mh\neq 0$ manifestly fails. The result of  
\cite{barb}, which was based on such approximation, 
becomes therefore meaningless,  and no hierarchy 
among the first and the second term of the asymptotic expansion should
 be expected. This is illustrated  in Fig.\ref{drho}.
Moreover, in the way they are compared in Fig. \ref{drho} and in Table 1, 
no reducible contribution induced by resummation of one-loop effects~\cite{chj}
 is included in  the $O(g^4 \mt^4)$ term.
Such separation between irreducible and reducible terms
 is not  possible for the second term of the HTE,
that, as we have seen, depends very strongly on the scheme.  This tells us that
reducible contributions (products of one-loop integrals) are 
very important there. But we know that the HTE
works very  well at one-loop level and that the leading quadratic
term is dominant, so there is some indication 
(no proof) that the first two terms of the
HTE should give a reasonable  approximation.
Concerning the irreducible contributions, 
the two-loop calculation of \cite{dgv,dgs,dg} is based on two-point
functions only~\cite{dg}. Unlike the case of three and four
point-functions~\cite{box}, the HTE seems to work quite well for self-energies,
as has been demonstrated in the case of QCD corrections in \cite{QCD3b} up to
three loops.

The preceding heuristic arguments are certainly not sufficient.
A first direct
test of the HTE for the two-loop \ew corrections 
 can be obtained by comparing the results
of \cite{dgv,dgs} with the calculation of \cite{weiglein}.
Bauberger and Weiglein (BW) calculated in \cite{weiglein} the two-loop
self-energies contributing to $\Delta r$ (i.e. to the prediction of $\mw$)
which contain the Higgs boson together with fermions through a
direct numerical evaluation of Feynman diagrams, which did not involve any
heavy mass expansion. This subset of diagrams is gauge invariant but not
ultraviolet finite. It cannot therefore be used to calculate observables
at fixed $\mh$, but gives information on the $\mh$ slope. 
Even on the slope, however, the information from BW 
is not complete at $O(g^4)$, since 
purely bosonic contributions as well as boxes and vertices containing the Higgs
boson are neglected. This approximation is
difficult to justify as the latter are not suppressed by any factor
wrt the fermion-Higgs two-point functions: on the contrary, 
if one considers large $\mh$ values $\sim 1$ TeV,
 as BW do, the purely bosonic contributions contain $\mh^2/\mw^2$ and 
$\ln^2 \mh/\mw\sim 5$
enhancement factors that may be dangerous to neglect. 
Moreover, at the one-loop level the  genuinely e.w. 
contributions to $\Delta r$ that are not enhanced by 
powers of $\mt$ are due to light fermions ($\approx 5\ 10^{-3}$), to subleading
terms from 
top-bottom loops ($\approx -5\ 10^{-3}$), and to bosonic loops, which reach
1.3\% for $\mh=1$ TeV. 
Keeping the first two and neglecting the last one is clearly 
not a good approximation at one-loop, and 
it is therefore unlikely to be so at the two-loop level.

\renewcommand{\arraystretch}{1.16}
\begin{table}
\begin{center}
\begin{tabular}{|c||c|c|c|c|} 
\hline 
 $\mh$  & 
 $ \Delta r_{(2),subtr}^{\rm OSII}$& $\Delta r^{\rm BW}_{(2),subtr} $ & diff & 
        $\delta\mw$  \\
(GeV)  & $(10^{-4})$ &  
$(10^{-4})$
&  $(10^{-4})$ &  ( MeV)\\ \hline 
100 &    -0.73 & -1.01 & -0.28 &0.4 \\ \hline
300 &    -3.10 & -3.97 & -0.87 &1.4     \\ \hline
600 &    -5.69 & -6.63 & -0.94 &1.5\\ \hline
1000  &  -9.44 & -10.45 &-1.01& 1.6 \\ \hline
\end{tabular} 
\caption{Comparison of the top-bottom contributions to the two-loop correction $\Delta r_{(2),subtr}$
from the calculations of \protect\cite{dgs} (OSII scheme) and BW 
\protect\cite{weiglein}.
QCD corrections are not included, and $\mw=80.37$ GeV is employed in the
evaluation of the radiative corrections. The second  and third columns give the
two-loop result from Refs. \protect\cite{dgs,weiglein}, 
respectively. The last column gives the differences that are induced in $\mw$. 
}
\end{center}\label{comp} 
\end{table} 
Despite these potential faults, 
the work of BW is an important step towards the
goal of a complete two-loop calculation. From our point of view,
moreover, it allows a nice partial check of the  HTE for what concerns the
$\mh$ slope,\cite{gsw} because BW did not use any mass expansion.
Indeed, one can isolate the top-bottom contributions from BW and compare 
them 
with the results of the OSII scheme of \ \cite{dgs}, after expanding $\Delta r$
in the numerator in order to follow the procedure of BW. 
 For simplicity, 
 we computed radiative corrections at a 
fixed $\mw=80.37$ GeV and removed all QCD corrections. The subtraction point
can be chosen at $\mh=65$ GeV, defining
 $\Delta r^{subtr}(\mh) =  \Delta r(\mh)-
\Delta r(65)$. The results are shown in Table 5.
We observe that the projected discrepancies for $\mw$ are very small, well
within the theoretical error reported in Table 4, even in the completely
unrealistic case of $\mh=1$ TeV, and that the maximum difference in the size of
the two-loop correction is less than 10\%. 
The HTE seems therefore to work quite well also at the two-loop level, at least
for what concerns the diagrams containing top and Higgs.  

From the above considerations it should be clear that discrepancies in the
calculation of   $\Delta r^{subtr}(1
{\rm TeV})$ are  unlikely to provide a good estimate of the overall 
theoretical error at more realistic values of $\mh$, where the $\chi^2$
distribution of the fits is centered. However,  a consistency check is
possible. It consists in considering all the two-loop contributions
calculated in \cite{dgs} and \cite{weiglein}, including also the light
fermions-Higgs   contributions of BW, without which their result would not be
well-defined. In that case, we find~\cite{gsw} a maximum discrepancy in the
calculation of $ \Delta r^{subtr}(\mh)$ at $\mh=1$ TeV, corresponding to 
$\delta\mw=1.7$ MeV. If one takes out from the calculation of \cite{dgs} 
the square of the one-loop bosonic contribution (a term neglected by BW), 
one finds an additional $-2.5$ MeV contribution, bringing the total difference
to $\delta\mw=-0.8$ MeV. 
A more comprehensive comparison of the HTE with 
 the work of BW (also  $\seff$ and $\Gamma_l$ 
 have been calculated in the same way \cite{weiglein2})  will be presented 
  in \cite{gsw}. 
Again, we conclude that 
deviations from \cite{dgs} appear to be within the range of Table 4, and that
the purely bosonic terms should not  be neglected if one wants to go beyond the
HTE: 
real improvement on the HTE awaits a complete two-loop calculation.

We now move on to the fit of the Higgs boson mass.
Understanding the main features 
of the global fit to $\mh$ can be facilitated by the use of 
simple formulas~\cite{dgps} that summarize the precise calculation 
of \cite{dgs}.
In the $\msbar$ scheme with $\alpha_s(\mz)=0.118$ and expressing $\mt$, $\mw$,
and $\mh$ in GeV and $\Gamma_l$ in MeV, we find
\bea
\frac{\sin^2\theta_{eff}^{lept}}{0.23151}= 1
+ 0.00226 \ln \frac{\mh}{100} + 0.0426 \left( \frac{\Delta\alpha_h}{0.028}-1
\right)\!-0.012 \!\left(\frac{\mt^2}{175^2}-1\right)
\label{seff}\eea
\bea
\frac{\mw}{80.383}&=& 1
- 0.00072\, \ln \frac{\mh}{100} - 1.0 \
10^{-4}\,\ln^2 \frac{\mh}{100}\ \ \ \nonumber\\
&&\ \ \ \ -0.00643 \left( \frac{\Delta\alpha_h}{0.028}-1\right)+
0.00676 \left(\frac{\mt^2}{175^2}-1\right)
\label{mw}\eea
\bea
\frac{\Gamma_l}{84.013}&=& 1
- 0.00064 \,\ln \frac{\mh}{100} - 0.00026\,\ln^2 \frac{\mh}{100}\ \ \ 
\nonumber\\
&&\ \ \ \ -0.00567 \left( \frac{\Delta\alpha_h}{0.028}-1\right)+
0.00954 {\left(\frac{\mt^2}{175^2}-1\right)}
\label{gammal}\eea
These formulas are very  accurate within  1$\sigma$ from the central values of
their inputs:
$170\lsim\mt\lsim 181$ GeV, $0.0273\lsim\Delta\alpha_h\lsim
0.0287$, 
and  for  $75 \lsim\mh\lsim 350$ GeV. In this range they reproduce 
the exact results of the calculation with maximal errors of 
$\delta s^2_{eff}\sim 1\times 10^{-5}$,
$\delta\mw\lsim 1 $MeV and  $\delta\Gamma_l\lsim 3$KeV, which are all very much
below the experimental accuracy. More complete expressions for $\seff$ and
$\mw$, including also the $\as$ dependence, can be found in \cite{dgps}.

By comparing the coefficients of $\ln \mh$ in \eqs{seff} and 
(\ref{mw}), we see that $\seff$ is 3
times more sensitive to $\ln \mh$ than $\mw$, 6.6 times more sensitive
to $\Delta\alpha_h$, almost 2 times more sensitive to $\mt$ and $\as$.
Despite the recent progresses in the measurement of $\mw$,
 it is therefore clear that most of the present
sensitivity to $\mh$ still comes from the effective sine.
Indeed, 
the world average \cite{teubert}, $\seff=0.23157\pm 0.00018$, can be used alone
to obtain an upper  bound on $\mh$ roughly comparable to the one of the global
fit. Using $\mt=(173.8\pm5)$ GeV and 
combining in quadrature the errors on $\mt$, 
$\seff$, $\Delta\alpha_h$ (the
conservative value of ~\cite{jeger}), one finds from \equ{seff} 
$\ln \mh/100 = 0.042\pm0.638$, which corresponds to $\mh=104^{+93}_{-49}$ GeV
or $\mh<297$ GeV at 95\% C.L. 
The theoretical error can be included in this estimate as a systematic error.
In fact, this  simple exercise can  be repeated in three different
schemes as done in \cite{dgps}, 
the respective central values can be averaged, and the error expanded
to cover the range of the three calculations. This gives $\mh<300$ GeV.
The QCD uncertainty can then be taken into account using the estimate discussed
above, $\delta \seff\approx \pm 3\times 10^{-5}$. This shifts the upper bound
by about 6\%, leading to $\mh<318$ GeV at 95\% C.L., 
which can be compared to the  global fit 
result~\cite{teubert} of $\mh<262$ GeV.
Like the EWWG, I am not taking into account the existence of 
a direct lower bound on $\mh$ from LEP~\cite{teubert}, $\mh> 89.8$ GeV,
which can be thought to play a role 
in deriving the $\mh$ fit.\cite{erler,chanowitz}

The very high sensitivity of $\seff$ to the inputs has its disadvantages,
however. In particular, this observable depends very strongly on the precise
value of the electromagnetic coupling at the $Z^0$ scale. We have seen in Table
2 that indeed  the present error on $\Delta\alpha_h$ may shift $\seff$ by 
2.3 $10^{-4}$, more than the error associated to the world average.
Even taking into account the new estimates of $\Delta\alpha_h$\cite{newdeltaa},
or future  low-energy measurements of $R_h$~\cite{jegertalk}, this
factor constitutes a major limitation of the resolving power on $\mh$.
In addition, the experimental situation for the $\seff $ measurement, although
better than a year ago, is far from satisfactory, given the unresolved 
discrepancy between LEP and SLD asymmetries.

\begin{figure}[t]
\centerline{
\mbox{%
\epsfig{file=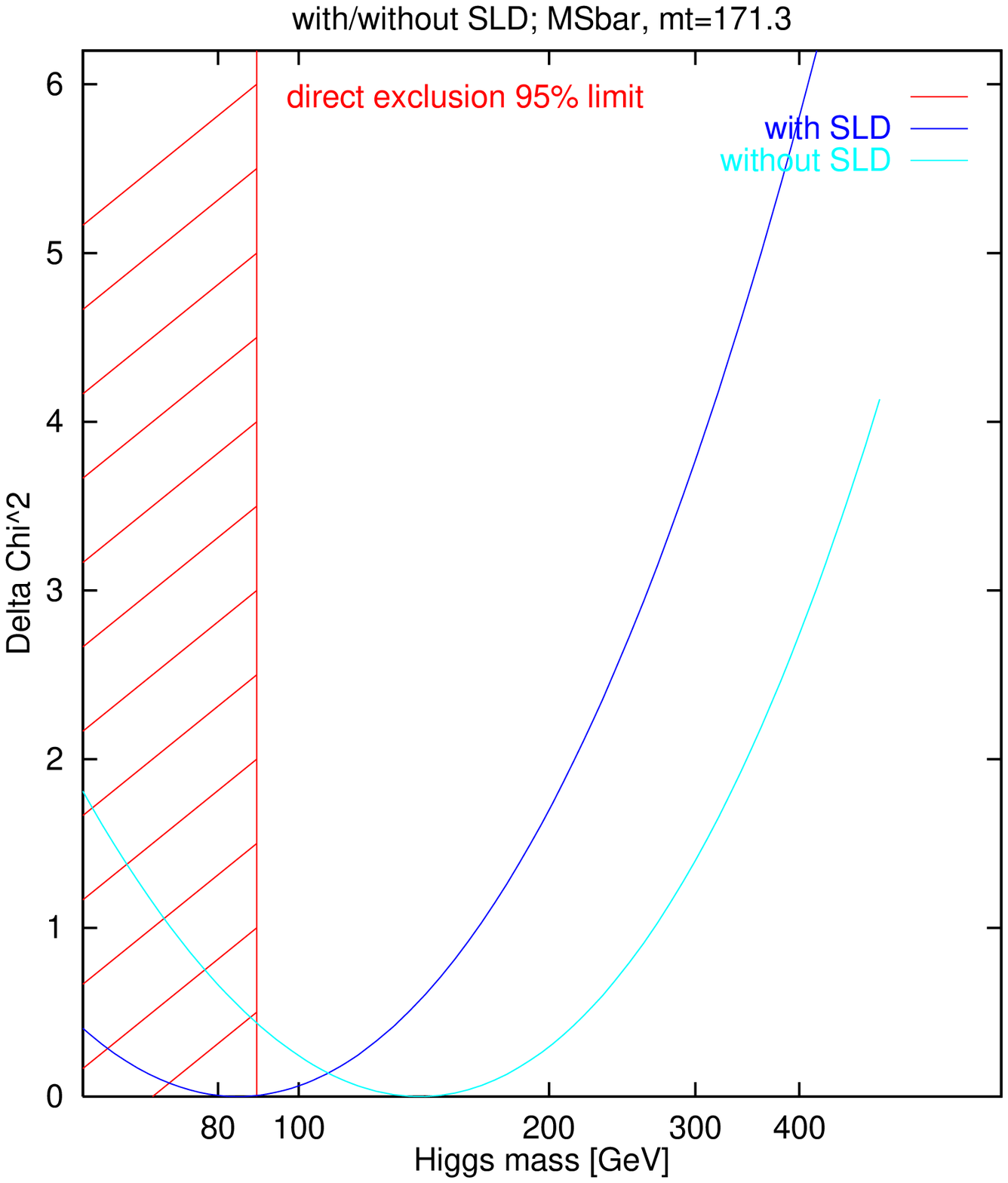 
        ,height=5.5cm  
        ,width=6.5cm   
}}
\hspace{-6mm}\mbox{%
\epsfig{file=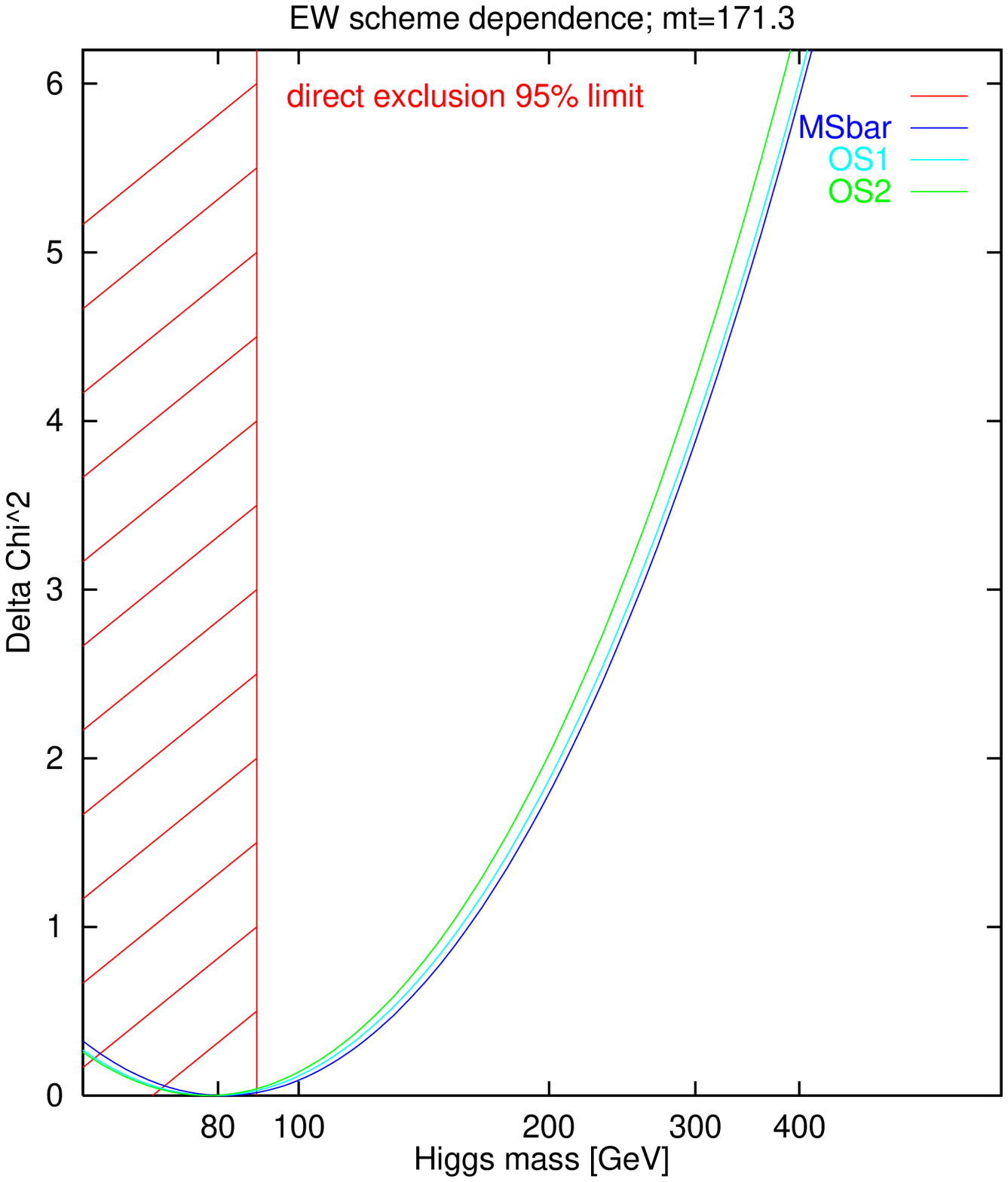 
        ,height=5.5cm  
        ,width=6.5cm   
       }%
}
}
\caption{$\mh$ fit based on $\seff$, $\mw$, $\Gamma_l$, $\mt=171.3$ GeV and
$\Delta\alpha_h=0.0280$, which reproduces the main features of the global 
fit~\protect\cite{teubert}.} 
\label{fit1}
\end{figure}
The measurement of $\mw$, on the other hand, can be considered complementary
to the one of the effective sine. At present, it easy to see from \equ{mw}
that,  if we try to determine $\mh$ from $\mw$ alone, $\delta x\equiv
\delta \ln \mh/100 \approx 1.05$. The $W$ mass has started to play a role in
the global fit to $\mh$, but is still far from competing with 
the effective sine, for which $\delta x=0.638$ (notice that 
$e^{1.05}\sim 2.9$ and that $e^{0.64}\sim 1.9$ are the relevant quantities).
However, assuming 
an error of 35 MeV on $\mw$  and of 2.5 GeV on $\mt$,
 and that the measurement of $\seff$ will   not
improve significantly in the next few years, one finds that $\delta x\approx
0.55$ for both the effective sine and for $\mw$ ($e^{0.55}\sim1.7$). 
This seems  to be a quite reasonable scenario for the near future,
 as the LEP200 experiments are still going to improve the $\mw$
 measurement~\cite{lancon},  
and the Run II at Tevatron, expected to start next year, should decrease
significantly the error on the top mass.\cite{tuts}
One can therefore conclude  (see also \cite{peppe})
that in a few years time the measurement of 
$\mw$ will provide the same sensitivity to the Higgs mass of the effective
sine, allowing an important check.

The essential features of the global fit to $\mh$ can be easily reproduced
using the three most precise measurements ($\seff$, $\mw$, and $\Gamma_l$)
and Eqs.(\ref{seff}-\ref{gammal}). Because of the strong correlation between
$\mt$ and $\mh$, apparent in Eqs.(\ref{seff}-\ref{gammal}), also observables 
insensitive to the Higgs boson have an indirect effect on the
$\mh$ fit. $R_b$, in particular, still points to a much lighter top 
($\mt=151\pm 25$ GeV) than most other data. We can take this effect into
account by using a lower $\mt=171.3$  instead of 173.8 GeV as input.
The $\mh$ fit obtained is very close to the global one and is shown in the
first plot of Fig. \ref{fit1}. Without  
including the theoretical errors, 
the 95\% C.L. upper bound on
$\mh$ is about 235 GeV and its central value 84 GeV.
 The exclusion of the SLD result for $\seff$ 
from the world average gives~\cite{teubert} $\seff=0.23189\pm0.00024$, which
leads to
$\mh\lsim 385$ GeV. I am not arguing here in favor of this exclusion. 
This last result simply shows  that even with a value of $\seff$ 
1.5$\sigma$  higher 
there would  be strong indication  for a light Higgs boson.  
\begin{figure}
\centerline{
\mbox{%
\epsfig{file=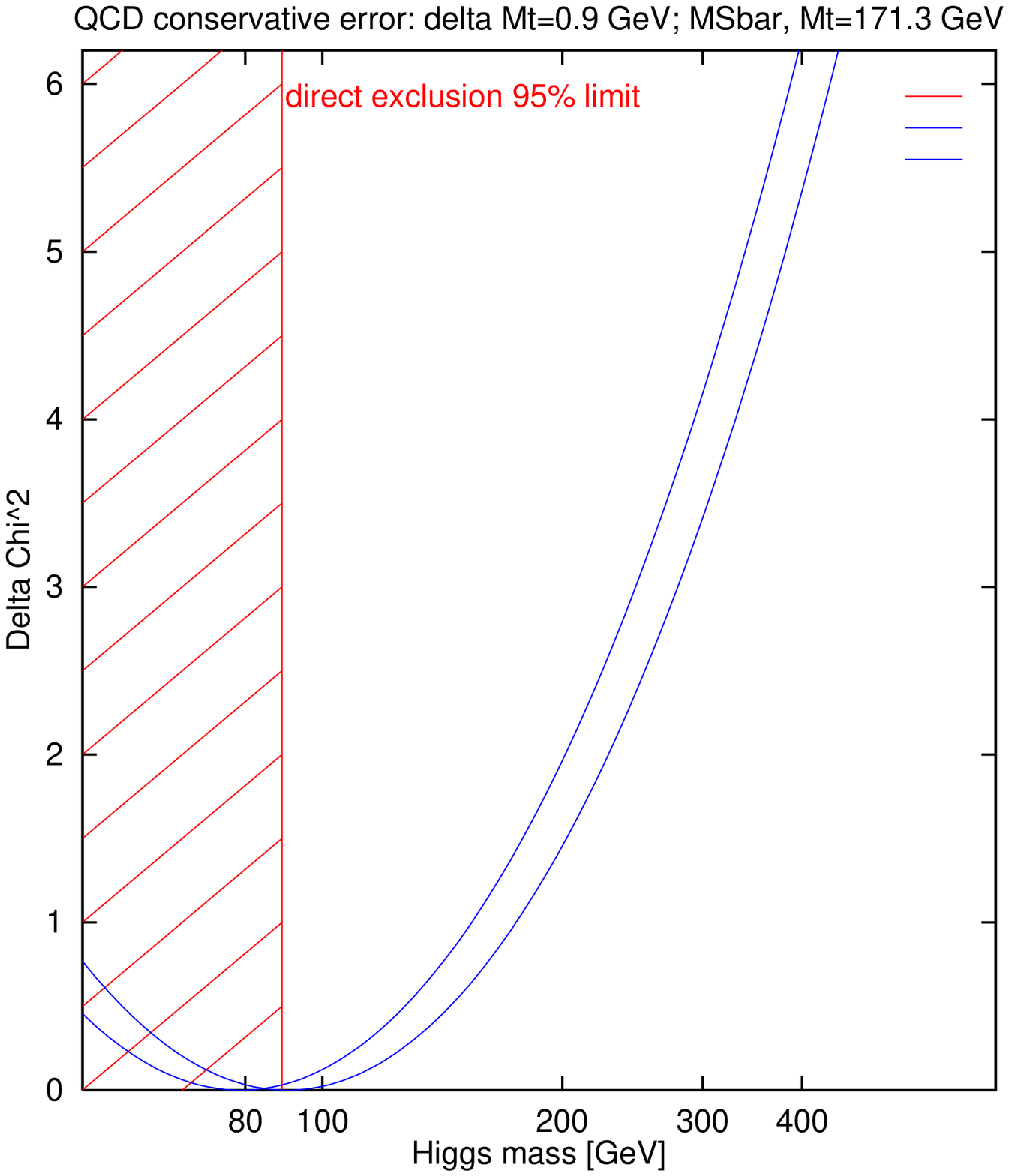 
        ,height=5.5cm  
        ,width=6.5cm   
}}
\hspace{-6mm}\mbox{%
\epsfig{file=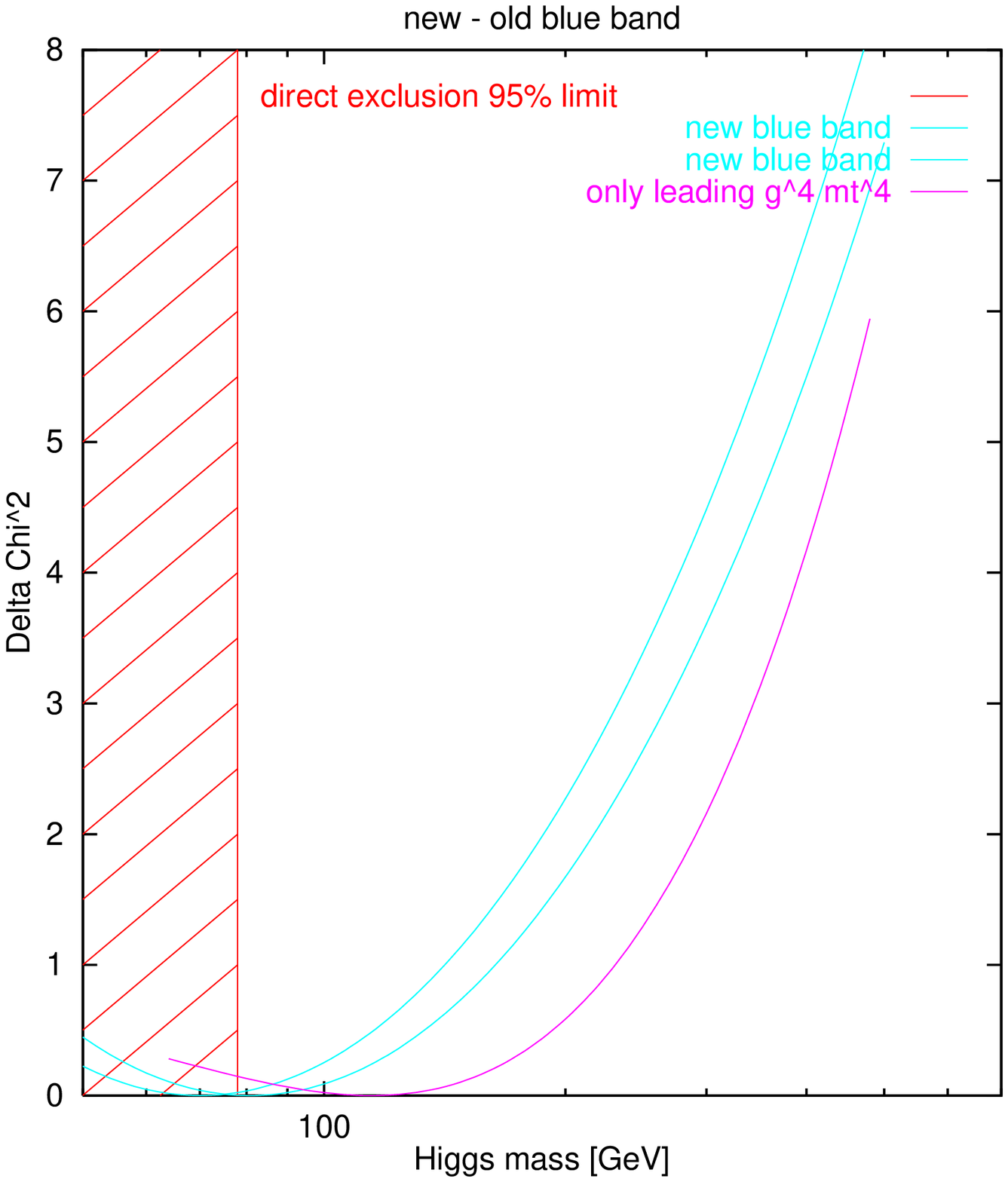 
        ,height=5.5cm  
        ,width=6.5cm   
       }%
}
}
\caption{As in Fig.\ref{fit1}, but considering the QCD uncertainty of the fit 
(plot on the left) and the combination of QCD and \ew uncertainties 
(plot on the right). }\label{fit2}
\end{figure}

The effect of various theoretical errors on the $\mh$ fit is shown in the
second plot of Fig.\ref{fit1} and in Fig.\ref{fit2}. We see that the \ew scheme
dependence has a very small effect on the fit. More important is the effect 
of the QCD uncertainty. Considering that 
most of it is linked to the leading $O(G_\mu \mt^2)$ 
contribution to $\Delta\rho$,
and in particular to the top quark mass definition,
we can implement it as a simple shift of $\mt$. The conservative 
values in Table 4 correspond to  a systematic shift
$\delta \mt=\pm 0.9$ GeV, displayed in the first plot of Fig. 4. 
The effect on the present fit is then an increase 
of about $+15$ GeV of the upper $\mh$ bound. The QCD and \ew uncertainties are 
combined in the second plot of Fig.\ref{fit2}, forming the analogue of the blue
band in Fig.1, with which there is  very good agreement. The upper bound on
$\mh$ is now 260 GeV.
The  result  of the same analysis
carried out without implementing the $O(g^4 \mt^2/\mw^2)$ corrections is also
considered.  It is clear that in that 
case the central value and  upper bound of $\mh$ are significantly
 larger, about 30 and 90 GeV, respectively.
\begin{figure}[t]
\centerline{
\mbox{%
\epsfig{file=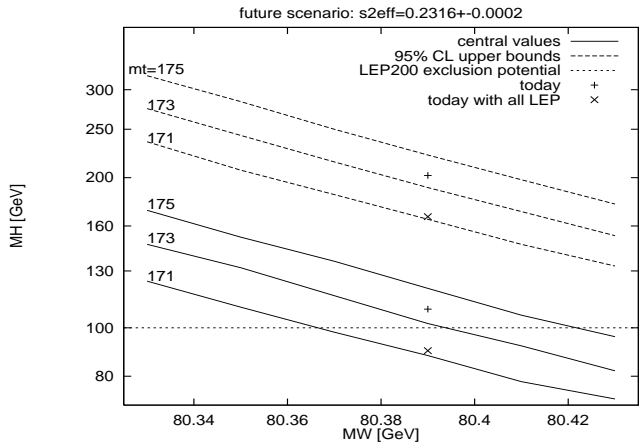 
        ,height=6cm  
        ,width=8cm   
}}}
\caption{Future scenarios for the indirect  determination of $\mh$, assuming
no change in the measurements of $\seff$  and in $\Gamma_l$, and $\delta
\mw=30$ MeV, 
 $\delta \mt=2.5$ GeV. The horizontal line marks the exclusion potential
of LEP200, the dashed lines the 95\% C.L. upper bounds on $\mh$ 
 and the solid lines the $\mh$ central values for a given
($\mw,\mt$) central value.
} \label{fut}
\end{figure}

Finally, the future scenario considered above  is 
analyzed in Fig.\ref{fut}, where the central
value and 95\% C.L. upper bounds on $\mh$ are reported as a function of the
central value of the measurement of $\mw$ for different values of $\mt$, under
the assumptions that i) the measurements of $\seff$ and $\Gamma_l$ 
will not change significantly; ii) the errors on $\mw$ and $\mt$ will 
decrease to 30 MeV and 2.5 GeV, respectively. The value
$\Delta\alpha_h=0.0278\pm 0.0003$, corresponding to the conservative scenario
of the second of Ref. \cite{newdeltaa}, has been used and the 
intrinsic theoretical errors
neglected. The conclusion is that under these two assumptions 
a determination of $\mh$ within about 80\% with a confidence level of  95\% 
  will be possible.
The results if the central values of $\mw$ and $\mt$ should not change
are  also marked, for the two cases $\mt=173.8$ and 171.3 GeV.  

\vspace{.3cm}
I am grateful to the organizers and especially to Joan Sola for organizing
this very interesting and 
 pleasant conference,  to D. Bardin, G. Degrassi, W. Hollik, G. Passarino, and
G. Weiglein for useful discussions and communications, 
and to A. Sirlin for reading the manuscript.

\end{document}